\documentclass[prb,twocolumn,showpacs,preprintnumbers]{revtex4}

\usepackage{graphicx,array,multirow,tabularx}
\newcolumntype{C}[1]{>{\centering\let\newline\\\arraybackslash\hspace{0pt}}m{#1}}

\begin{document}
\preprint{}

\title{Termination-dependent Electronic and Magnetic Properties \\
of Ultrathin SrRuO$_{3}$ (111) Film on SrTiO$_{3}$}

\author {Bongjae Kim and B. I. Min}
\affiliation{Department of Physics, PCTP,
Pohang University of Science and Technology, Pohang 790-784, Korea}
\date{\today}

\begin{abstract}

 We have investigated electronic and magnetic properties of ultrathin
SrRuO$_{3}$ (SRO) film grown on (111) SrTiO$_{3}$ substrate
using the {\it ab initio} electronic structure calculations.
Ru-terminated SRO (111) film suffers from strong surface atomic relaxations,
while SrO$_{3}$-terminated one preserves the surface structure of ideal perovskites.
Both Ru- and SrO$_{3}$-terminated SRO (111) film
show unexpected interlayer antiferromagnetic (AFM) structure at the surface,
but with different characters and mechanisms.
The AFM structure for the former results from the large surface atomic relaxation,
whereas that for the latter results from the truncated film effect.
Interestingly, for the SrO$_{3}$-termination case, the half-metallic nature
emerges despite the interlayer AFM structure.
Upon reducing the thickness, the collapsing behavior of magnetic anisotropy from
out-of-plane to in-plane easy axis is found to occur for the Ru-termination case,
which, however, does not pertain to SrO$_{3}$-termination case.

\end{abstract}


\maketitle


\section{Introduction}

 Recent advances in thin-film growth technique make it possible to achieve delicate control of
oxide interfaces in the atomic scale, so as to bring
an extra dimension to the physics of transition-metal oxide systems.
LaAlO$_{3}$-SrTiO$_{3}$ (LAO-STO) heterostructure is a typical example,
which exhibits either metallic or insulating feature depending on the selection of
interface type.\cite{ohtomo,nakagawa}
 In magnetic heterostructures, even more fascinating features can be obtained
by utilizing systematic control of relevant material
parameters.\cite{may,dong1,bhattacharya,freeland,che,gibert}

SrRuO$_{3}$ (SRO), which is the only ferromagnetic (FM) metal in perovskites,
has been a subject of intense study in its film form since it can be well-grown
epitaxially as a single-crystal.\cite{koster}
 In the ultrathin SRO film grown on SrTiO$_{3}$ along (001) direction, the metal-insulator
transition (MIT) was observed with decreasing the thickness, which was accompanied
by sudden collapse of magnetic anisotropy and vanishing of $T_{c}$.\cite{toyota,chang1,xia}
Establishment of antiferromagnetic (AFM) order and/or quantum confinement (QC) were
thought to be the origin of these critical behaviors.\cite{mahadevan,xia,chang1}
There are several issues currently in dispute for SRO (001) film.
On the critical thickness of SRO film for the MIT,
there is a discrepancy in existing reports;
3$-$4 monolayers (MLs) of SRO unit cell~\cite{toyota,xia}
vs. 1$-$2 MLs.\cite{chang1}
The character of the magnetism is another unresolved issue,
{\it i.e.}, whether SRO thin film belongs to a system of
itinerant or local magnetism.\cite{park,hdkim,maiti,rondinelli,etz,jeong,grebinskij,Shai13}
Even the basic understanding of electronic and magnetic properties of ultrathin SRO film,
whether it is metallic or insulating, FM or AFM, and so on, are far from complete.
For instance, the half-metallic nature at the interface of SRO/STO superlattices
was predicted,\cite{alves} but it has not been experimentally verified yet.

\begin{figure}[b]
\includegraphics[angle=0,width=65mm]{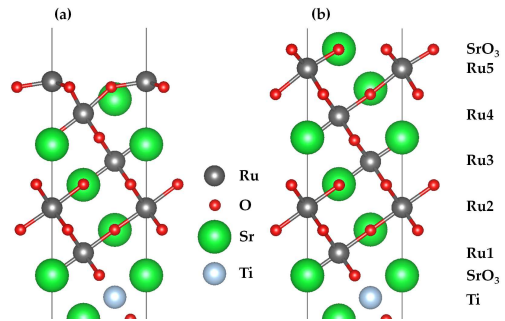}
\caption{(Color online)
Five MLs of SRO (111) film grown on top of the STO substrate.
(a) Ru-termination, and (b) SrO$_{3}$-termination.
Sr, Ru, O and Ti atoms are represented by green, gray, red,
and light blue balls, respectively.
}
\label{structure}
\end{figure}

 In contrast to numerous reports on SRO (001) film, the exploration on SRO (111) film has been
relatively rare. In fact, systematic growth of SRO (111) film
on STO was accomplished only recently.\cite{chang3,grutter1,bwlee10,grutter2}
In the experimental aspect, delicate control
of the layer growth is required, especially for the first ML, due to the conversion tendency of
highly volatile Ru$^{4+}$ to SrO$_{3}^{4-}$.\cite{chang3,rijnders}
It is naturally expected that changing the growth direction
of system would produce
totally distinct physical properties due to different coordination environment.
Indeed, recent studies on
other perovskite (111) film or heterostructure
systems showed rich exotic phenomena.\cite{may,gibert,eom,paudel,herranz,doennig,xiao,yang}
 As shown in Fig.~\ref{structure}, SRO along the (111) direction is comprised of alternating
Ru$^{4+}$ and SrO$_{3}^{4-}$ layers. Thus there are two different termination type:
Ru- and SrO$_{3}$- termination.

Intriguingly, from the $M$-$H$ curve of SRO (111) films,
Grutter \emph{et al.}\cite{grutter1} obtained the magnetization
as large as 3.4 $\mu_{B}$/Ru,
which far exceeds the low-spin (LS) moment value of 2 $\mu_{B}$/Ru.
So they claimed the stabilization of the high-spin (HS) state (4$\mu_{B}$/Ru)
in the SRO (111) films.
Also, from the x-ray magnetic circular dichroism (XMCD) spectroscopy,
they deduced the ratio of orbital/spin magnetic moment to obtain
the large orbital magnetic moment of 0.24-0.32 $\mu_{B}$/Ru.\cite{grutter2}
However, detailed study of electronic structure and magnetic properties of SRO (111)
films is lacking,
and moreover, the HS phase of SRO (111) film has not been reproduced by other groups yet.

In this paper, we have investigated characteristic features of ultrathin SRO (111) film
grown on STO, and compared its properties to those of ultrathin SRO (001) film.
 Based on the result of {\it ab initio} electronic structure calculations,
we have predicted termination-dependent magnetic behaviors of ultrathin SRO (111) film.
Surprisingly, we have found that the interlayer AFM structures
are stabilized at the surfaces for both Ru- and SrO$_{3}$-termination cases,
but through fundamentally different mechanisms.
Further, we have obtained the half-metallic property for the
SrO$_{3}$-terminated SRO (111) film despite its surface interlayer
AFM structure.
As for the spin state, we have not found any evidence of HS state
in SRO (111) films, contrary to the report by Grutter \emph{et al.}.\cite{grutter1}

\section{Methodology}

 The calculations were performed by using the plane-wave basis set and the projector-augmented-wave
method implemented in the Vienna \emph{ab initio} simulation package (VASP).\cite{vasp}
Additional calculations using the full-potential linearized augmented plane-wave (FLAPW) band method
were also employed for detailed analysis.\cite{freeman,wien2k}
 For the exchange-correlation energy functional, we used the generalized gradient approximation (GGA)
with PBEsol functional.\cite{perdew1} On-site Coulomb correlation $U$ is treated with the GGA$+U$ method
in the rotationally invariant form \cite{dudarev} and the spin-orbit coupling (SOC) was taken into account
when necessary.

Firstly, lattice parameter of the bulk STO substrate
is optimized and compared with the experimental value. In-plane lattice parameter is obtained to be 5.51{\AA},
which is less than 0.2$\%$ difference compared to the bulk value of 5.52{\AA}.
On top of optimized six layers of STO unit cell, SRO film with fixed lattice parameter to substrate STO is
systematically deposited up to five MLs for both terminations. Cases of 5MLs of SRO film (111) are shown
in Fig. ~\ref{structure}. On top of SRO film, 15{\AA} vacuum was considered in all film calculations.

Plane-wave energy cutoff of 500 eV was used with Monkhorst-Pack mesh \emph{k}-point sampling of $7\times7\times1$.
Overall calculations are checked up to $11\times11\times1$ grid.
For the determination of the magnetic structure, we have considered every possible magnetic structures and full
relaxations of atoms are done strictly until the force of each atom is below 0.1 meV/{\AA} for each case.
When considering the intralayer AFM structure, we doubled the in-plane
unit cell and similar criteria were applied.
For the calculation of the magnetic anisotropy energy (MAE),
we used the fully relaxed structures. With the inclusion
of SOC, the difference between in-plane and out-of-plane energies were calculated
and the \emph{k}-point convergence was checked up to $13\times13\times1$ with the VASP code.

 As mentioned earlier, the correlation effect in SRO system has not been settled yet.
Nevertheless, recent detailed calculations found that, for SRO systems,
the weak correlation limit is
apparently better to describe the experimental findings such as
photoemission~\cite{rondinelli} and Curie temperature.~\cite{etz}
For $U>3$ eV, the DFT+$U$ (DFT: density functional theory) gives the half-metallic phase,
which is not consistent
with experiment, and even for $U=2$ eV, the calculated magnetic moment already exceeds
the highest reported experimental value.\cite{granas}
Therefore, in view of the existing results,
we adopted the weak correlation value of $U_{eff}$ = 1.0 eV,
which describes the experimental results well.
The relation of magnetic structure and correlation will be discussed
further in the section III. C.

\section{Results and discussions}

 Magnetic properties of SRO film are strongly correlated with its RuO$_{6}$ tilting
and rotations.\cite{singh,middey} Hence the layer-dependent structural evolution
is expected to play an important role in the magnetic structure.\cite{chang2, jhe, gu}
 In the case of SRO (001) film, the degree of layer-dependent octahedral distortion
is still controversial, but the suppression of tilting and
rotation due to the substrate is not perfect even
in the ultrathin limit.\cite{jhe,chang2}
Therefore, the magnetic structure of ultrathin (001) film depends not only on the number of MLs but also on
the evolution of tilting patterns that changes Ru-O-Ru bond connectivities.
 In the case of (111) growth, however, only the trigonal distortion is active
because of the symmetry nature.
This feature distinguishes (111) from (001) and (110) cases, for which tetragonal and monoclinic
distortion occur, respectively.\cite{grutter1}
Accordingly, for the description of physics of SRO (111) ultrathin film,
bulk tilting or rotational degree of freedom does not need to be considered.

\subsection{Ru-termination case}

\begin{figure}[b]
\includegraphics[angle=0,width=85mm]{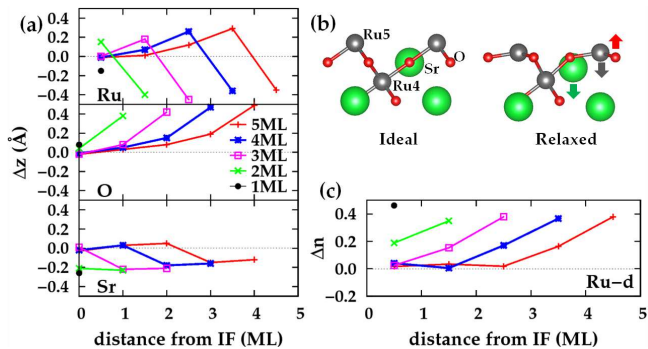}
\caption{(Color online)
Atomic displacements and occupations of Ru-terminated SRO (111) films
for different ML numbers.
(a) Vertical displacements $\Delta$z ($\AA$) of Ru, O, and Sr ions
with respect to the ideal positions.
The position of each ion is set to be zero for the ideal case.
(b) Atomic structures at the surface for ideal and relaxed cases.
(c) Relative occupation difference ($\Delta$n) with respect to that of bulk SRO.
}
\label{Rmove}
\end{figure}

Let us discuss the Ru-terminated SRO (111) film first.
We have obtained that the ideal SRO (111) film without the surface atomic relaxation
has a simple FM structure, as in bulk SRO, regardless of ML numbers.
The surface Ru atom has a highly enhanced magnetic moment of 2.6$-$2.7$\mu_{B}$/Ru
with respect to those of inner Ru atoms (1.4$-$1.5$\mu_{B}$) (see Table~\ref{moments}).
This is attributed to the exposure of magnetic ion with reduced screening effects.
 For the relaxed SRO (111) film, however, completely different behaviors are found.
Atomic relaxations near the surface are very large
so as to give the strongly altered crystal field to the surface Ru atom.
As shown in Fig.~\ref{Rmove}(a), due to the small compressive strain from
STO substrate, SRO blocks slightly expand along the growth direction, which is shown
by general outward movements of inner atoms. At the surface, however,
the movement patterns are distinguished from those of inner layers.
Most notable is that the surface Ru and O atoms, respectively, move in and out greatly,
and so eventually almost planar RuO-like layer is formed at the surface,
as described schematically in Fig.~\ref{Rmove}(b).
It is seen in Fig.~\ref{Rmove}(a)
that the subsurface layer shows the relaxation patterns
in-between those of inner and surface layers.

\begin{table}[t]
\centering
\caption{Layer dependent magnetic moments ($\mu_{B}$) at Ru sites.
Ru5 corresponds to the surface Ru atom.
}
\begin{ruledtabular}
\begin{tabular}{C{0.8cm}C{1.2cm}|r r|r r}
\multirow{2}{*}{}   &         &\multicolumn{2}{c|}{~~~Ru-termination~~~}& \multicolumn{2}{c}{SrO$_{3}$-termination}\\
                 ML & Ru type &ideal&relaxed&ideal&relaxed\\
\hline\hline
     1 & 1 & 0.0 & 0.0 ~~~& 1.2 & 1.2 ~~~\\
\hline
  \multirow{2}{*}{2} & 1 & 1.5 & 0.3 ~~~&  1.4 &  1.6 ~~~\\
                     & 2 & 2.7 & 1.0 ~~~& $-$1.2 & $-$1.5 ~~~\\
\hline
  \multirow{3}{*}{3} & 1 & 1.4 & 1.3 ~~~&  1.3 &  1.4 ~~~\\
                     & 2 & 1.6 & 0.8 ~~~&  1.3 &  1.5 ~~~\\
                     & 3 & 2.6 & 0.8 ~~~& $-$1.2 & $-$1.5 ~~~\\
\hline
  \multirow{4}{*}{4} & 1 & 1.4 &  1.2 ~~~&  1.4 & 1.3 ~~~\\
                     & 2 & 1.5 &  1.3 ~~~&  1.4 & 1.2 ~~~\\
                     & 3 & 1.6 & $-$0.3 ~~~&  1.4 & $-$1.6 ~~~\\
                     & 4 & 2.6 & $-$0.9 ~~~& $-$1.2 & 1.6 ~~~\\
\hline
  \multirow{5}{*}{5} & 1 & 1.4 &  1.2 ~~~&  1.3 & 1.4 ~~~\\
                     & 2 & 1.4 &  1.3 ~~~&  1.4 & 1.5 ~~~\\
                     & 3 & 1.5 &  1.2 ~~~&  1.4 & 1.4 ~~~\\
                     & 4 & 1.5 &  0.0 ~~~&  1.3 & 1.5 ~~~\\
                     & 5 & 2.6 & $-$0.8 ~~~& $-$1.2 & $-$1.6 ~~~\\

\hline\hline
  \multicolumn{2}{c|}{bulk}&\multicolumn{4}{c}{1.3}\\
\end{tabular}
\end{ruledtabular}
\label{moments}
\end{table}

\begin{figure}[b]
\includegraphics[angle=270,width=85mm]{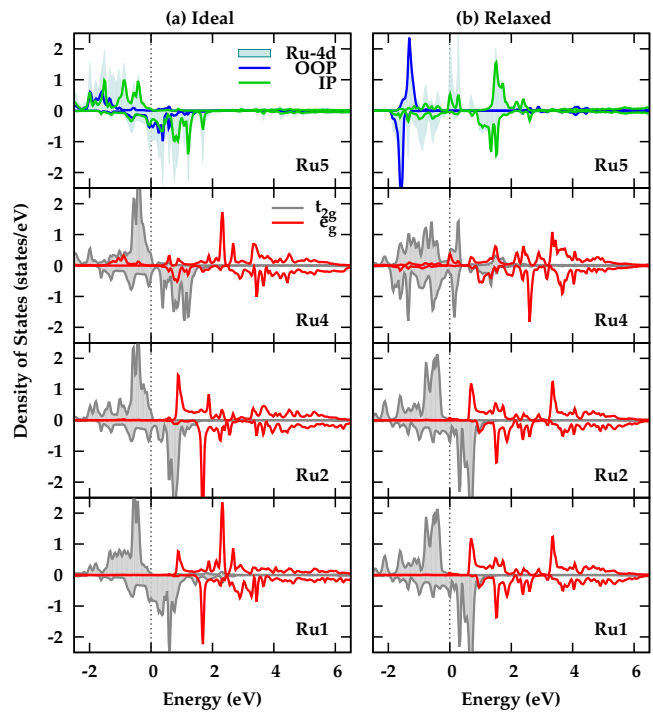}
\caption{(Color online)
Ru partial DOS (PDOS) of (a) ideal and (b) relaxed SRO (111) film ($N=5$)
for Ru-terminated case.
PDOS of Ru3 is not shown.
}
\label{Rdos}
\end{figure}

The substantial atomic rearrangements near the surface brings about
dramatic modification of electronic and magnetic structures.
Figure~\ref{Rdos} provides the Ru-$d$ partial density of states (DOS)
of both ideal and relaxed SRO (111) film ($N=5$) in its most stable magnetic structure.
Interestingly, the ideal SRO (111) film has FM half-metallic nature, as shown in
Fig.~\ref{Rdos}(a).
We have found that the half-metallic nature is sustained for all $N$
in the ideal case.
Ru5 at the surface is seen to have the HS state.
 After relaxation, the spins of inner Ru ions (Ru1, Ru2, and Ru3)
are ferromagnetically aligned, while the spin of surface Ru ion (Ru2) is
to be antiferromagnetically coupled to those of inner Ru ions.
As we can see in Fig.~\ref{Rdos}(b) of the relaxed case,
the out-of-plane (OOP) $d$-states of Ru1 are fully occupied,
whereas the in-plane (IP) $d$-states
are almost empty, the pattern of which is quite different from that of ideal structure case.
This makes contrast to DOSs of inner Ru ions (Ru1, Ru2, and Ru3) that
have normal DOS forms in the octahedral crystal field with four
$d$ electrons occupying low-lying $t_{2g}$ states.
The DOS of subsurface Ru (Ru4) is also changed considerably
compared to those of inner layers,
implying that the subsurface acts as a boundary between the surface and bulk layers.
It is seen in Fig.~\ref{Rdos}(b) that the half-metallic nature is
retained for inner layers of Ru1-Ru3, but disappears for
surface layers of Ru5 and Ru4 as the interlayer AFM structure evolves at the surface.

Due to substantial Ru-$d$ orbital rearrangements
for surface Ru5 and Ru4 ions,
the occupancies are also changed significantly.
Namely the surface Ru atom has larger electron occupation compared to inner Ru
atoms by as much as 0.4 electrons (Figure~\ref{Rmove}(c)).
Here the occupation numbers were obtained from the total charges
in the Wigner-Seitz cell of the atoms.
These occupation numbers generally do not give the exact valences,
but the difference of them can be thought as the valence electron change.
These features will be reflected in the charge densities too.
Planar RuO$_{3}$ polyhedra at the surface would have prominent
out-of-plane charge density shapes,
which are highly deformed from those of inner almost-ideal octahedral RuO$_{6}$.

\begin{figure}[t]
\includegraphics[angle=0,width=85mm]{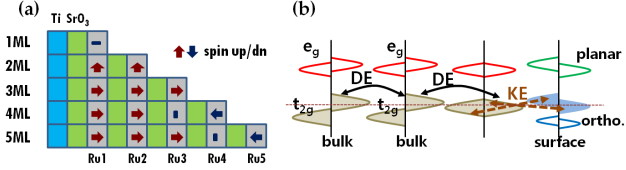}
\caption{(Color online)
The case for Ru-terminated SRO (111) film.
(a) Magnetic structures for different ML number $N$.
(b) Schematic magnetic interactions near the surface.
The DE represents the double exchange that corresponds to the magnetic interaction
arising from the first order in direct hopping ($\sim t_{eff}$),
while the KE represents the kinetic exchange that corresponds to that
arising from the second order in indirect hopping ($\sim t_{eff}^2$).
}
\label{Rmag}
\end{figure}

In view of the stable FM structure for the ideal SRO (111) film,
the strong surface atomic relaxation for the (111) Ru-terminated case
is suggestive of the the driving mechanism of its surface
interlayer AFM structure.
In the case of bulk SRO, the double exchange (DE) mechanism is known
to be responsible for its itinerant ferromagnetism.\cite{etz}
But, for the SRO (111) film case,
the electron hopping strength between bulk and surface Ru ions becomes weakened
due to the substantial rearrangements of Ru-$d$ states near the surface.
Thereby, the interlayer AFM structure follows as
the kinetic exchange (KE) mechanism wins over the DE at the surface,
as provided schematically in Fig.~\ref{Rmag}(b).

 Thus the competition between the FM DE and AFM KE interactions
determines the magnetic structures of
the Ru-terminated SRO (111) films. As shown in Fig.~\ref{Rmag}(a), up to $N=3$,
the DE seems to win over the KE to have the stable FM structure.
For $N=3$, the FM phase is energetically more stable
than the surface interlayer AFM phase by
3.7 meV/Ru. But, for $N$ $>$ 3, the KE wins over the DE at the surface
to have the interlayer AFM phase (surface layer itself is FM).
For $N$=4 and 5, the surface interlayer AFM phases are lower in energy
than the FM phases by 11.1 meV/Ru and 13.1 meV/Ru, respectively.
This feature suggests that $N$=3 can be regarded as
the critical ML number for the change of the magnetic structure.
For the (111) Ru-terminated  case, only the $N=1$ SRO film, which simply
corresponds to the Ru monolayer on top of the STO block, has the nonmagnetic and
insulating ground state.

\subsection{SrO$_{3}$-termination case}

\begin{figure}[t]
\includegraphics[angle=0,width=80mm]{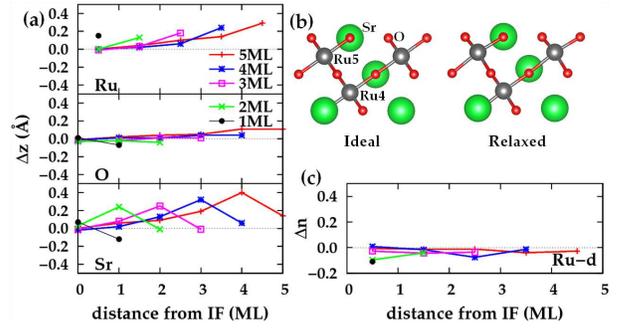}
\caption{(Color online)
Atomic displacements and occupations of SrO$_{3}$-terminated SRO (111) films
for different ML numbers.
(a) Vertical displacements $\Delta$z ($\AA$) of Ru, O, and Sr ions
with respect to the ideal positions.
The position of each ion is set to be zero for the ideal case.
(b) Atomic structures at the surface for ideal and relaxed cases.
(c) Relative occupation difference ($\Delta$n) with respect to that of bulk SRO.
}
\label{Smove}
\end{figure}

\begin{figure}[b]
\includegraphics[angle=270,width=85mm]{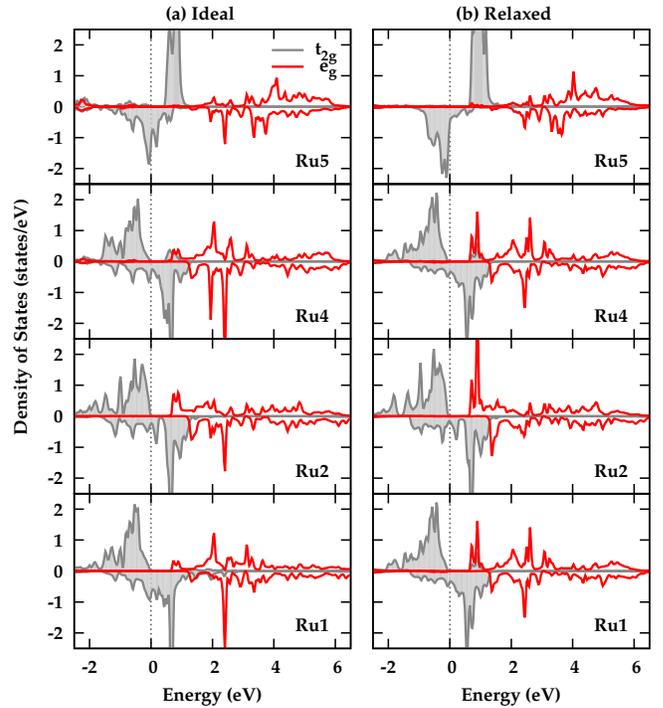}
\caption{(Color online)
Ru partial DOS (PDOS) of (a) ideal and (b) relaxed SRO (111) film ($N=5$)
for SrO$_{3}$-terminated case. PDOS of Ru3 is not shown.
}
\label{Sdos}
\end{figure}

\begin{figure}[b]
\includegraphics[angle=0,width=85mm]{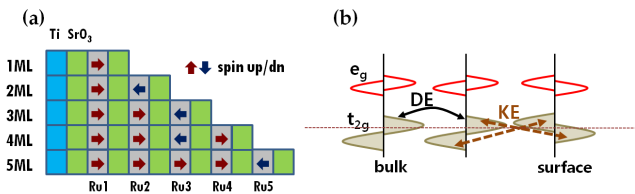}
\caption{(Color online)
The case for SrO$_{3}$-terminated SRO (111) film.
(a) Magnetic structures for different ML number $N$.
(b) Schematic magnetic interactions near the surface.
}
\label{Smag}
\end{figure}

Electronic and magnetic properties of the SrO$_{3}$-termination case are
fundamentally different from those of Ru-termination case.
Firstly, the surface atomic movements are not so prominent as for the Ru-termination case.
As shown in Fig.~\ref{Smove}(a) and (b), except for the surface Sr atoms
that tend to avoid exposure by moving in, the atomic relaxation gives simple
outward expansions of the SRO blocks due to lattice mismatch of the SRO and STO.
Hence, as we can see in Fig.~\ref{Sdos}, the overall shapes of DOSs are similar
between the ideal
and the relaxed cases, and Ru-$d$ occupation at the surface is also almost the same as
those of inner layers (Fig.~\ref{Smove}(c)).
Surprisingly, the interlayer AFM phase at the surface emerges
for SrO$_{3}$-termination case too.
Distinctly from the Ru-termination case,
the AFM phase occurs even from $N$=2 (see Fig.~\ref{Smag}(a)).\cite{4LS}
Furthermore, the surface interlayer AFM structure is
already stabilized for an ideal film without the relaxation, which suggests that
the atomic relaxation is not an essential ingredient for the surface magnetic structure
of the SrO$_{3}$-termination case.
Since the relaxation at the surface is not considerable, even after the relaxation,
the connectivity of Ru-O-Ru bond, which plays a key role in
the magnetic interactions, is hardly changed from the ideal case.
One can thus deduce that the driving mechanism of surface interlayer AFM structure for the
SrO$_{3}$-termination case would be different from that for the Ru-termination case.

As mentioned above, in the case of (111) growth, RuO$_{6}$ tilting and
rotations are suppressed by symmetry,
and so the Ru-O-Ru bond angle remains almost straight for every layers.
This symmetry driven bond character
plays a crucial role in determining magnetic interactions
at the surface of SrO$_{3}$-terminated (111) film.
Note that the DE interaction is proportional to the effective
Ru-Ru hopping $t_{eff}$, while the KE to its square.
Hence the latter is more affected by the Ru-O-Ru bond angle
than the former.
As a result, the straight Ru-O-Ru bond would favor the AFM KE
over the FM DE interaction.
Indeed it was reported that the FM phase in the cubic phase of bulk SrRuO$_{3}$
is stabilized only after the consideration of the longer range
DE interaction.\cite{etz}

 For SrO$_{3}$-terminated (111) film,
farther neighbor interactions are prohibited at the surface
due to truncated film structure.
Hence the AFM KE interaction becomes dominant over the FM DE interaction
at the surface, and so the interlayer AFM structure can be stabilized,
as shown in Fig.~\ref{Smag}(b).
The DOS in Fig.~\ref{Sdos}(b) reveals that only the
spin direction is flipped over at the surface Ru atom with respect to those of inner Ru atoms
without much change in the shape of DOS.
Very similar DOS shapes for ideal and relaxed cases unambiguously represent
the crucial role of truncated structure in the surface interlayer AFM structure.
Noteworthy is that SRO (111) film for the SrO$_{3}$-termination case has
the half-metallic nature even for the relaxed case of Fig.~\ref{Sdos}(b).
We have ascertained that this feature of surface interlayer AFM and
half-metallic nature is sustained for every ML cases.

\subsection{Magnetic anisotropy, spin-state, and intralayer magnetic structure}

\begin{figure}[b]
\includegraphics[angle=270,width=85mm]{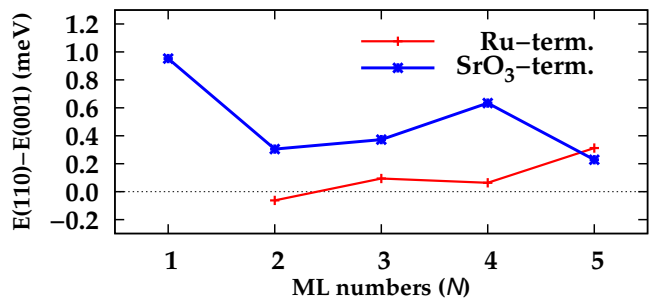}
\caption{(Color online)
Calculated MAE of both Ru- and SrO$_{3}$-terminated SRO (111) films
for different ML numbers.
For Ru-termination case, the sudden collapse to the in-plane anisotropy
occurs for $N=2$, while, for SrO$_{3}$-termination case,
the out-of-plane anisotropy persists down to $N=1$.
Energy difference is calculated per Ru ion.
}
\label{mae}
\end{figure}

Recent magneto-optic Kerr experiment on ultrathin SRO (001) film detected the
collapse of magnetic anisotropy from OOP to IP at around
3 MLs.\cite{xia}
In the case of SRO (111) film, the OOP direction was reported to be
the easy axis,\cite{grutter2,bwlee14} but the magnetic anisotropy in the ultrathin limit
has not been explored yet.
Figure~\ref{mae} provides the calculated MAE for both termination cases.
As shown in Fig.~\ref{Rmag}(a) for the Ru-termination case,
the OOP anisotropy is stable down to $N$=3, and then
sudden collapse to the IP anisotropy occurs for $N$=2,
which is reminiscent of the behavior of the SRO (001) film.\cite{xia}
For the SrO$_{3}$-termination case in Fig.~\ref{Smag}(a), however,
the OOP anisotropy persists down to $N$=1.

Overall magnetic anisotropy patterns resemble the magnetic structure evolutions
in a sense that $N$=3 acts as a critical ML number for the Ru-termination case, while
no such critical behavior is seen for the SrO$_{3}$-termination case.
The MAEs between OOP and IP
directions are found to be very tiny, only of less than meV scale.
This feature is consistent with the obtained small orbital magnetic moment of less than
0.1 $\mu_{B}$/Ru for every cases.
Notewwothy is that this finding does not agree with the estimated orbital magnetic moment
of 0.24-0.32 $\mu_{B}$/Ru by Grutter \emph{et al.}\cite{grutter2}
from the XMCD experiment.

\begin{figure}[t]
\includegraphics[angle=270,width=85mm]{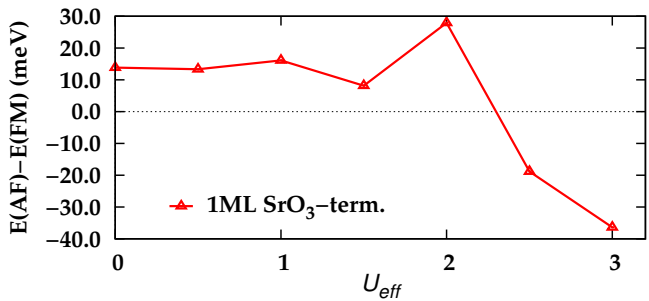}
\caption{(Color online)
Energy difference between intralayer AFM and FM structures
as a function of $U_{eff}$ parameter
for the 1 ML SrO$_{3}$-terminated SRO (111) film case.
}
\label{U}
\end{figure}

Regarding the spin-state of the Ru$^{4+}$,
Grutter \emph{et al.}\cite{grutter1,grutter2} reported the magnetization value as much
as 3.4 $\mu_{B}$/Ru from the $M$-$H$ data, and claimed that Ru in SRO (111) film has
the HS configuration.
Their claim, however, is hardly reconciled with our systematic calculations of magnetization
in the ultrathin SRO (111) film, which manifest that Ru in SRO (111) film is in the LS state
for both Ru- and SrO$_{3}$-termination cases.
Even though we impose the HS configuration (4 $\mu_{B}$) initially,
the LS phase is quickly stabilized.
The trigonal distortion, which was proposed to be the origin of
the HS configuration,
is not strong enough because the lattice mismatch between STO and SRO is not large
(overall expansion of one ML of SRO is less than 0.1{\AA}).
Moreover, a recent experiment on the same system reported the stable LS phase,\cite{bwlee14}
which contradicts to the report by Grutter \emph{et al.}.
We thus think that more detailed experimental analyses are demanded
for the clarification of the HS phase of Ru ion in SRO (111) film system.

Lastly, let us briefly discuss the correlation effect on the magnetic structures of
SRO (111) film.
We examined the stability of the surface intralayer (not interlayer)
AFM structure for the SRO (111) film.
We considered the 1 ML SrO$_{3}$-termination case,
which corresponds to the thinnest magnetic film.
As shown in Fig.\ref{U}, we found that
the FM structure is more stable than the intralayer AFM structure
as long as $U_{eff} \leq$ 2 eV.
This implies that the SRO (111) surface layer would have the
stable intralayer FM structure as long as it belongs to the itinerant weak correlated system.
Remember that experimental findings in SRO are better described
by the DFT calculation in the the weak correlation limit.
Hence this result demonstrates that our predicted  surface interlayer AFM structures for
SRO (111) films are robust.
On the other hand, for $U_{eff} >2$ eV, the the intralayer AFM structure becomes stabilized.
One comment here is that, for SRO (001) film,
Mahadevan \emph{et al.}\cite{mahadevan} reported that the intralayer AFM structure
is a ground state at the surface, which is the origin of the MIT.
They used somewhat larger correlation parameter of $U$=2.5eV with $J$=0.4eV.
However, for smaller $U=0$, they also obtained a stable FM structure.
Both ours and Mahadevan \emph{et al.}'s result indicate that
the use of correct correlation $U$ value is very important
to describe the physics of SRO systems properly.

\section{Conclusions}

We have studied electronic and magnetic structures of SRO film grown on (111) STO substrate using
the \emph{ab initio} band structure calculations.
We have found that the interlayer AFM structures emerge at the surfaces of both
Ru- and SrO$_{3}$-termination cases. For the Ru-termination case, the interlayer AFM comes from the
large atomic relaxation at the surface and the consequent dominant KE interaction.
For SrO$_{3}$-termination case, the growth symmetry prevents the tilting of Ru-O-Ru angles
to favor the AFM KE interaction over the FM DE interaction within the truncated film structure.
We have also found that, with decreasing the film thickness, the collapse of magnetic easy axis occurs
for the Ru-termination case, but not for SrO$_{3}$-termination case.
The half-metallicity is found to exist for the SrO$_{3}$-termination case of SRO (111) film,
but not for the Ru-termination case.
Our findings in ultrathin SRO (111) film, which are not accessible in SRO (001) film,
will provide important insight in tailoring new synthetic metal-oxide systems.

\acknowledgments

 We would like to thank B. H. Kim for helpful discussions.
This work was supported by the NRF (No.2009-0079947),
and by the KISTI supercomputing center (No. KSC-2012-C3-056).


\begin{thebibliography}{99}
   \bibitem{ohtomo}
    A. Ohtomo and H. Y. Hwang,
    Nature (London) {\bf 427}, 423 (2004).
   \bibitem{nakagawa}
    N. Nakagawa, H. Y. Hwang, and D. A. Muller,
    Nat. Mater. {\bf 5}, 204 (2006).
   \bibitem{may}
    S. J. May, P. J. Ryan, J. L. Robertson, J.-W. Kim, T. S. Santos, E. Karapetrova,
    J. L. Zarestky, X. Zhai, S. G. E. te Velthuis, J. N. Eckstein, S. D. Bader,
    and A. Bhattacharya,
    Nat. Mater. {\bf 8}, 892 (2009).
   \bibitem{dong1}
    S. Dong, R. Yu, S. Yunoki, G. Alvarez, J.-M. Liu, and E. Dagotto,
    Phys. Rev. B {\bf 78}, 201102(R) (2008).
   \bibitem{bhattacharya}
    A. Bhattacharya, S. J. May, S. G. E. te Velthuis, M. Warusawithana, X. Zhai,
    B. Jiang, J.-M. Zuo, M. R. Fitzsimmons, S. D. Bader, and J. N. Eckstein,
    Phys. Rev. Lett. {\bf 100}, 257203 (2008).
   \bibitem{freeland}
    J. W. Freeland, J. Chakhalian, A. V. Boris, J.-M. Tonnerre, J. J. Kavich,
    P. Yordanov, S. Grenier, P. Zschack, E. Karapetrova, P. Popovich, H. N. Lee,
    and B. Keimer,
    Phys. Rev. B {\bf 81}, 094414 (2010).
   \bibitem{che}
    C. He, A. J. Grutter, M. Gu, N. D. Browning, Y. Takamura, B. J. Kirby,
    J. A. Borchers, J. W. Kim, M. R. Fitzsimmons, X. Zhai, V. V. Mehta, F. J. Wong,
    and Y. Suzuki,
    Phys. Rev. Lett. {\bf 109}, 197202 (2012).
   \bibitem{gibert}
    M. Gibert, P. Zubko, R. Scherwitzl, J. \'{I}\~{n}iguez, and J.-M. Tricone,
    Nat. Mater. {\bf 11}, 195 (2012).
   \bibitem{koster}
    G. Koster, L. Klein, W. Siemons, G. Rijnders, J. S. Dodge, C.-B. Eom,
    D. H. A. Blank, and M. R. Beasley,
    Rev. Mod. Phys. {\bf 84}, 253 (2012)
    and references therein.
   \bibitem{toyota}
    D. Toyota, I. Ohkubo, H. Kumigashira, M. Oshima, T. Ohnishi M. Lippmaa,
    M. Takizawa, A. Fujimori, K. Ono, M. Kawasaki, and H. Hoinuma,
    Appl. Phys. Lett. {\bf 87}, 162508 (2005).
   \bibitem{chang1}
    Y. J. Chang, C. H. Kim, S.-H. Phark, Y. S. Kim, J. Yu, and T. W. Noh,
    Phys. Rev. Lett. {\bf 103}, 057201 (2009).
   \bibitem{xia}
    J. Xia, W. Siemons, G. Koster, M. R. Beasley, and A. Kapitulnik,
    Phys. Rev. B {\bf 79}, 140407(R) (2009).
   \bibitem{mahadevan}
    P. Mahadevan, F. Aryasetiawan, A. Janotti, and T. Sasaki,
    Phys. Rev. B {\bf 80}, 035106 (2009).
   \bibitem{park}
    J. Park, S.-J. Oh, J.-H. Park, D. M. Kim, and C.-B. Eom,
    Phys. Rev. B {\bf 69}, 085108 (2004).
   \bibitem{hdkim}
    H.-D. Kim, H.-J. Noh, K. H. Kim, and S.-J. Oh,
    Phys. Rev. Lett. {\bf 93}, 126404 (2004).
   \bibitem{maiti}
    K. Maiti, and R. S. Singh,
    Phys. Rev. B {\bf 71}, 161102(R) (2005).
   \bibitem{rondinelli}
    J. M. Rondinelli, N. M. Caffrey, S. Sanvito, and N. A. Spaldin,
    Phys. Rev. B {\bf 78}, 155107 (2008).
   \bibitem{etz}
    C. Etz, I. V. Maznichenko, D. B\"{o}ttcher, J. Henk, A. N. Yaresko, W. Hergert,
    I. I. Mazin, I. Mertig, and A. Ernst,
    Phys. Rev. B {\bf 86}, 064441 (2012).
   \bibitem{jeong}
   D. W. Jeong, H. C. Choi, C. H. Kim, S. H. Chang, C. H. Sohn, H. J. Park,
   T. D. Kang, D.-Y. Cho, S. H. Baek, C. B. Eom, J. H. Shim, J. Yu, K. W. Kim,
   S. J. Moon, and T. W. Noh,
    Phys. Rev. Lett. {\bf 110}, 247202 (2013).
   \bibitem{grebinskij}
    S. Grebinskij, \v{S}. Masys, S. Mickevic\v{i}us, V. Lisauskas, and V. Jonauskas,
    Phys. Rev. B {\bf 87}, 035106 (2013).
   \bibitem{Shai13}
   D. E. Shai, C. Adamo, D. W. Shen, C. M. Brooks, J. W. Harter, E. J. Monkman, B. Burganov,
   D. G. Schlom, and K. M. Shen,
   Phys. Rev. Lett. {\bf 110}, 087004 (2013).
   \bibitem{alves}
    M. Verissimo-Alves, P. Garc\'{i}a-Fern\'{a}ndez, D. I. Bilc, P. Ghosez,
    and J. Junquera,
    Phys. Rev. Lett. {\bf 108}, 107003 (2012).
   \bibitem{chang3}
    J. Chang, Y.-S. Park, J.-W. Lee and S.-K. Kim,
    J. Crys. Growth {\bf 311}, 3771 (2009).
   \bibitem{grutter1}
    A. J. Grutter, F. Wong, E. Arenholz, A. Vailionis, and Y. Suzuki,
    Appl. Phys. Lett. {\bf 96}, 082509 (2010).
   \bibitem{bwlee10}
    B. W. Lee and C. U. Jung, Appl. Phys. Lett. {\bf 96}, 102507 (2010).
   \bibitem{grutter2}
    A. J. Grutter, F. J. Wong, E. Arenholz, A. Vailionis, and Y. Suzuki,
    Phys. Rev. B {\bf 85}, 134429 (2012).
   \bibitem{rijnders}
    G. Rijnders, D. H. A. Blank, J. Choi, and C.-B. Eom,
    Appl. Phys. Lett. {\bf 84}, 505 (2004).
   \bibitem{eom}
    C.-B. Eom,
    APS March Meeting, G12.00005 (2013).
   \bibitem{paudel}
    T. Paudel, and E. Tsymbal,
    APS March Meeting, G12.00006 (2013).
   \bibitem{herranz}
    G. Herranz, F. Sanchez, N. Dix, M. Scigaj, and J. Fontcuberta,
    Sci. Rep. {\bf 2}, 758 (2012).
   \bibitem{doennig}
   D. Doennig, W. E. Pickett, and R. Pentcheva,
   Phys. Rev. Lett. {\bf 111}, 126804 (2013).
   \bibitem{xiao}
    D. Xiao, W. Zhu, Y. Ran, N. Nagaosa, and S. Okamoto,
    Nat. Commun. {\bf 2}, 389 (2011).
   \bibitem{yang}
   K.-Y. Yang, W. Zhu, D. Xiao, S. Okamoto, Z. Wang, and Y. Ran,
   Phys. Rev. B {\bf 84}, 201104(R) (2011).
   \bibitem{vasp}
    G. Kresse and J. Furthmuller, Phys. Rev. B {\bf 54}, 11169 (1996);
    G. Kresse and D. Joubert, \emph{ibid} {\bf 59}, 1758 (1999)
   \bibitem{freeman}
	H. J. F. Jansen and A. J. Freeman, Phys. Rev. B  {\bf 30}, 561 (1984).
   \bibitem{wien2k} P. Blaha {\it et al.}  \emph{WIEN2k,
	ISBN 3-9501031-1-1}. Karlheinz Schwarz, Techn. Universit\"{a}t Wien,
	Austria, 2001.
   \bibitem{perdew1}
    J. P. Perdew, A. Ruzsinszky, G. I. Csonka, O. A. Vydrov,
    G. E. Scuseria, L. A. Constantin, X. Zhou, K. Burke,
    Phys. Rev. Lett. {\bf 100}, 136406 (2008).
   \bibitem{dudarev}
    S. L. Dudarev, G. A. Botton, S. Y. Savrasov, C. J. Humphreys, and A. P. Sutton,
    Phys. Rev. B {\bf 57}, 1505 (1998).
   \bibitem{granas}
    O. Gr{\aa}n\"{a}s, I. D. Marco, O. Eriksson, L. Nordstr\"{o}m, and C. Etz, arXiv:1312.0270.
   \bibitem{singh}
    D. J. Singh, J. Appl. Phys. {\bf 79}, 4818 (1996).
   \bibitem{middey}
    S. Middey, P. Mahadevan, and D. D. Sarma,
    Phys. Rev. B {\bf 83}, 014416 (2011).
   \bibitem{chang2}
    S. H. Chang, Y. J. Chang, S. Y. Jang, D. W. Jeong, C. U. Jung, Y.-J. Kim,
    J.-S. Chung, and T. W. Noh,
    Phys. Rev. B {\bf 84}, 104101 (2011).
   \bibitem{jhe}
    J. He, A. Borisevich, S. V. Kalinin, S. J. Pennycook, and S. T. Pantelides,
    Phys. Rev. Lett. {\bf 105}, 227203 (2010).
   \bibitem{gu}
    M. Gu, Q. Xie, X. Shen, R. Xie, J. Wang, G. Tang, D. Wu, G. P. Zhang,
    and X. S. Wu,
    Phys. Rev. Lett. {\bf 109}, 157003 (2012).
   \bibitem{4LS}
The magnetic structure for $N=4$ looks different from other $N$ cases.
However, two surface interlayer AFM structures of
down-up and up-down spin configurations for Ru4-Ru3 are very close in energy,
(as small as 8.8 meV/Ru), compared to the FM spin configuration
that has much higher energy of more than 49 meV/Ru.
   \bibitem{bwlee14}
    B. Lee, O.-U. Kwon, R. H. Shin, W. Jo, and C. U. Jung,
    Nanoscale Res. Lett. {\bf 9}, 8 (2014).

\end{thebibliography}
\end{document}